\documentclass[manuscript,screen,nonacm]{acmart}

\copyrightyear{2024}

\begin{document}

\newcommand{\todo}[1]{{}}

\title{Generative AI for 2D Character Animation}

\author{Jaime Guajardo}
\author{Ozgun Bursalioglu}
\author{Dan B Goldman}
\affiliation{%
  \institution{Google}
  \country{USA}
}


\begin{abstract}
  In this pilot project, we teamed up with artists to develop new workflows for 2D animation while producing a short educational cartoon. We identified several workflows to streamline the animation process, bringing the artists' vision to the screen more effectively.
\end{abstract}

\begin{teaserfigure}
    \centering
    \includegraphics[height=24mm]{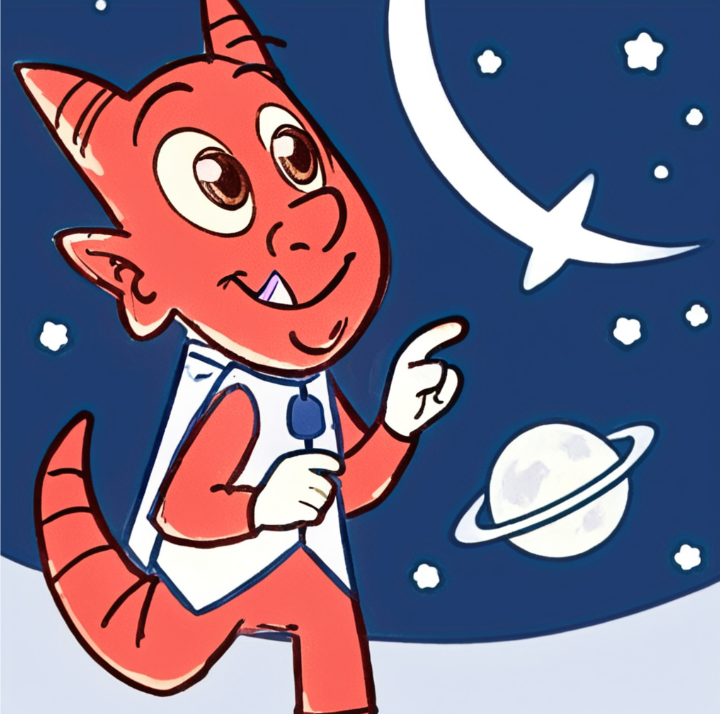}
    \includegraphics[height=24mm]{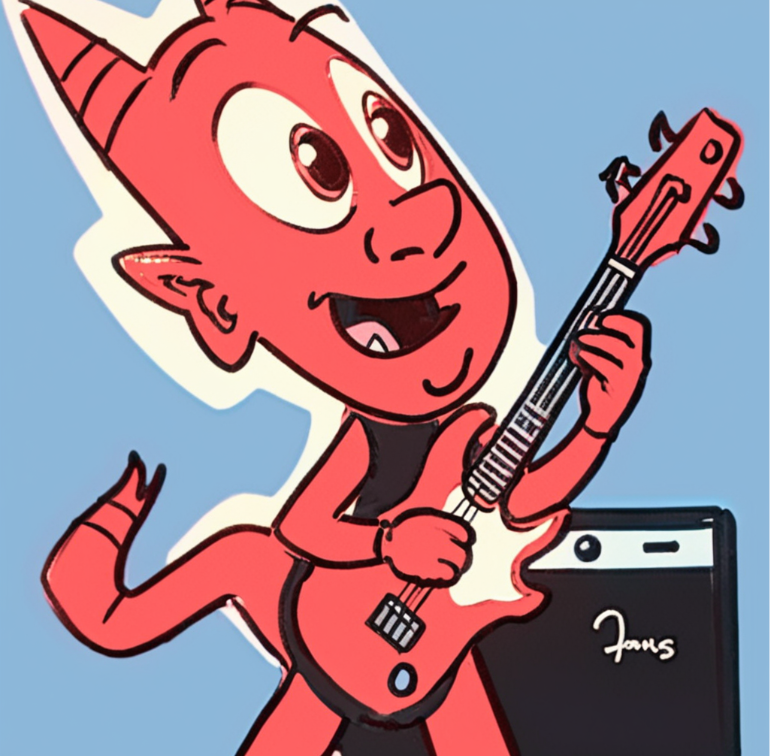}
    \includegraphics[height=24mm]{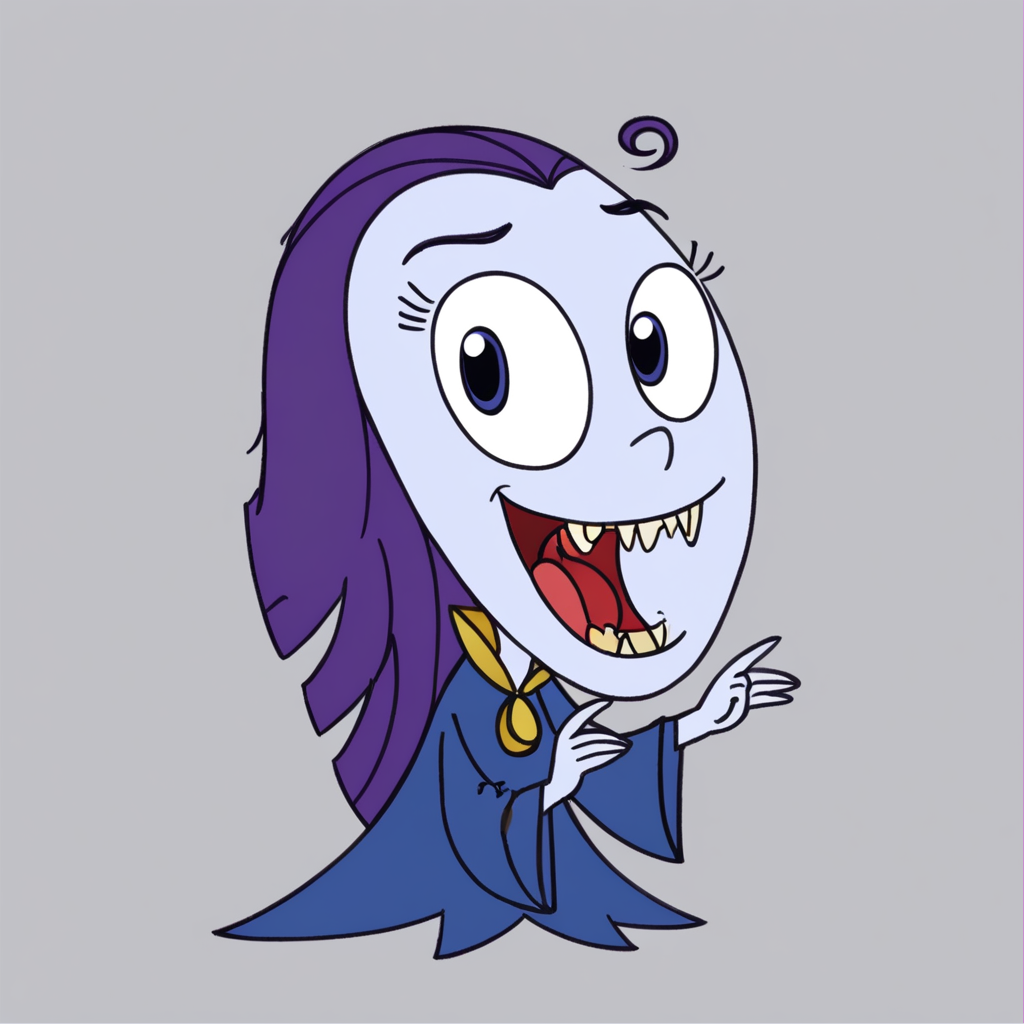}
    \includegraphics[height=24mm]{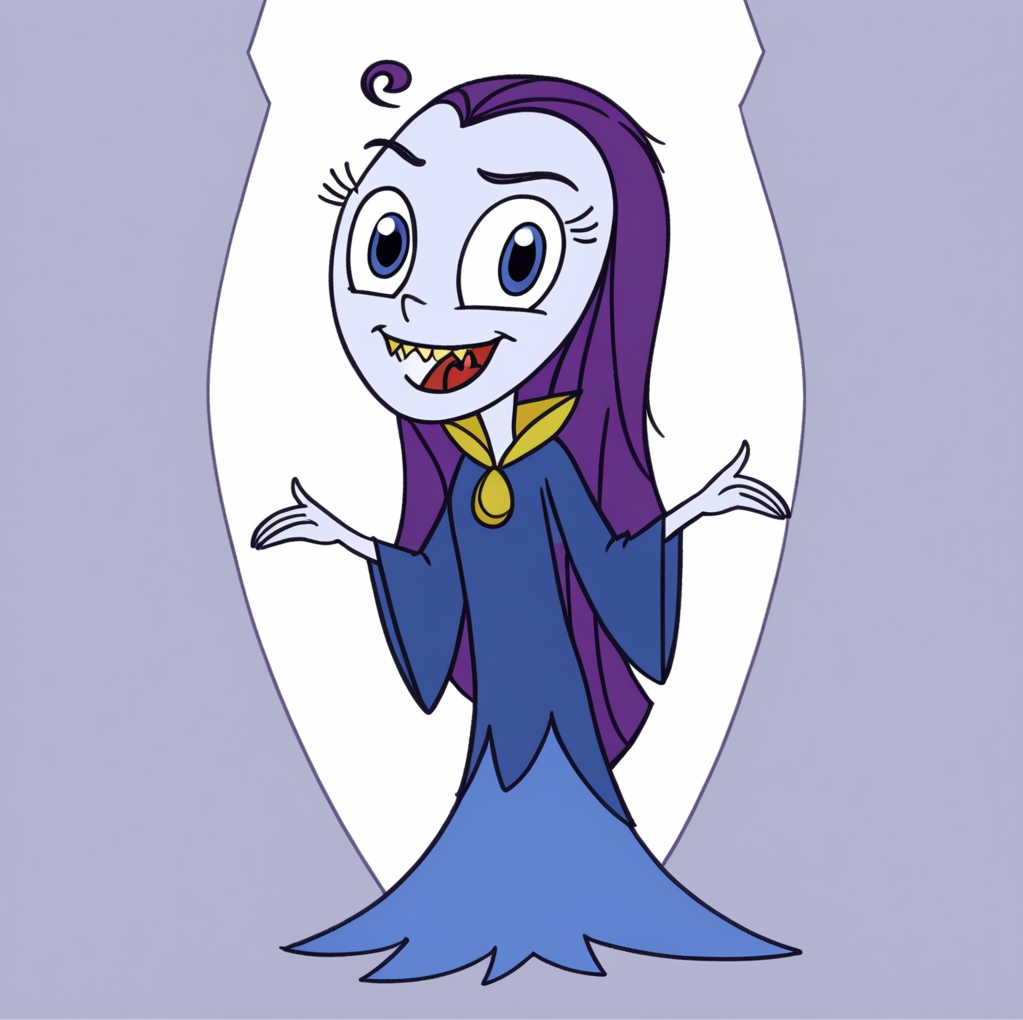}
    \includegraphics[height=24mm]{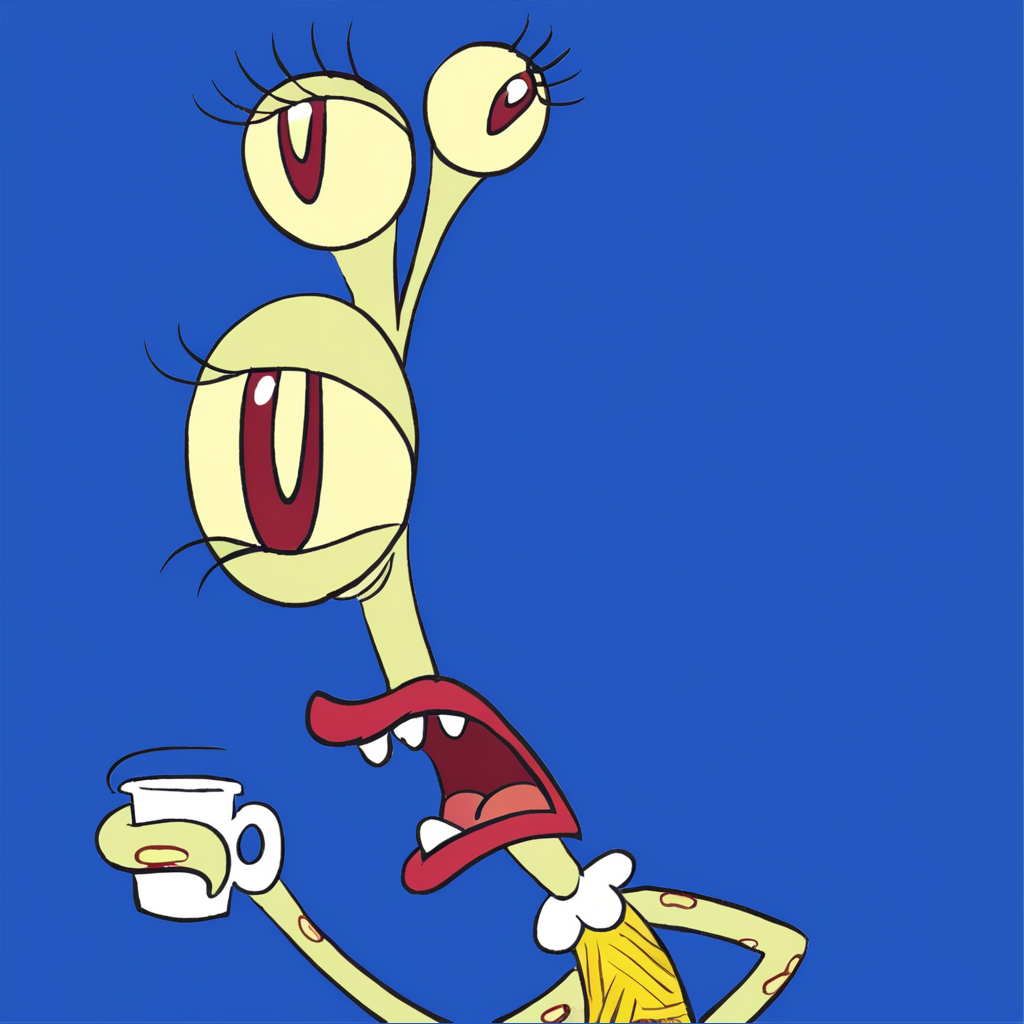}
    \includegraphics[height=24mm]{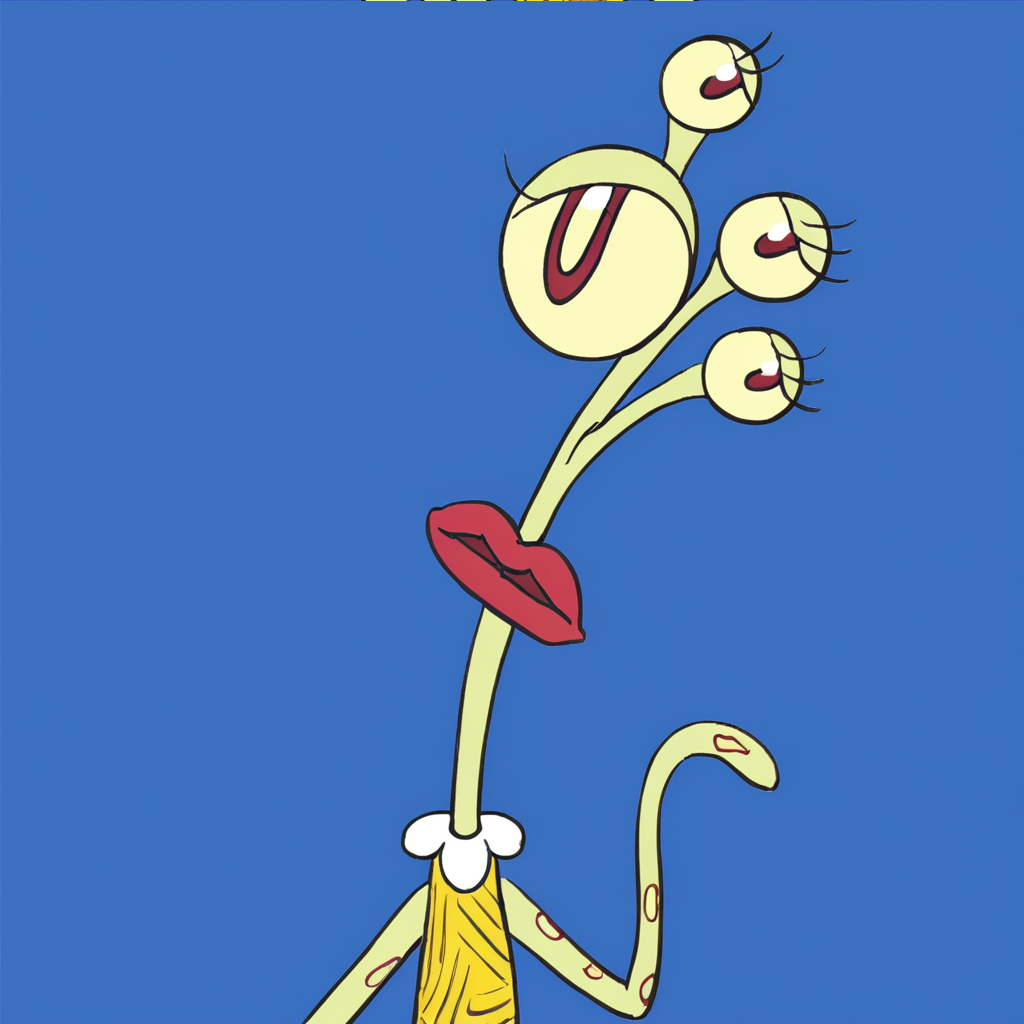}
    \Description{Text-to-image outputs from our Dreambooth SDXL character models for FRED, VERNA and MS. MENACE}
    \label{fig:teaser}
\end{teaserfigure}

\maketitle

\section{Introduction}

We produced a short educational cartoon using novel generative AI workflows. Educational videos are under-served relative to entertainment videos, and educators don't generally have the skill or talent to produce animations that teach real concepts. We adopted a 2D style because it can be more forgiving of the kind of geometric inconsistencies that generative AI sometimes produces, which limit the perceived quality of realistic or 3D-style imagery. Our narrative occurs in a fantasy world with some non-human characters, giving further flexibility for variable appearance.

\section{Workflows}

\begin{figure}[htb]
    \centering
    \includegraphics[width=0.66\linewidth]{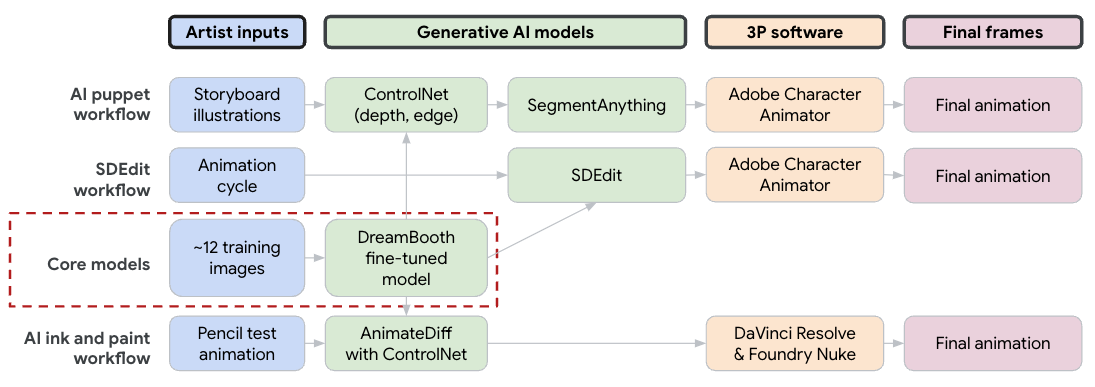}
    \caption{Schematic diagram showing three of our generative AI animation workflows}
    \Description{Schematic diagram showing three of our generative AI animation workflows}
    \label{fig:workflows}
\end{figure}

Previous generative animation projects~\cite{Showrunner} have focused on automating story simulation and visual design, but in this work, our goal is to accelerate the process of character animation. We identified several promising workflows (see Figure~\ref{fig:workflows}), bringing artists' vision to the screen more effectively.

\paragraph{SDXL fine tuning.} All our workflows begin with fine-tuned diffusion models. However, no available models have been trained extensively on the style of 2D cartoon characters that we designed. Furthermore, in early stages of production with a small crew, we had only a handful of illustrations for each character. We obtained best results by fine-tuning SDXL~\cite{podell2023sdxl} using DreamBooth~\cite{ruiz2023dreambooth}. Although we attempted to train LoRAs~\cite{hu2021lora} to avoid the need for full custom models for each character, the results were not as good. Even after full fine-tuning, the models did not always produce consistent proportions, line weights, or paint colors. Non-human characteristics --- like a tail or three eyes --- were especially problematic, often blending with other body parts (see teaser image). In most cases we were able to overcome these inconsistencies using artist drawings as additional conditioning, with ControlNet~\cite{zhang2023adding}. Although we tried both sketch and edge conditioning, we found that even though our characters are illustrated in a flat 2D style, depth conditioning worked best to match the specific pose of these input line drawings. Since SDXL produces only RGB images, each character's training images were first matted onto a contrasting background color. In this way the models more reliably output a constant background color for easy chroma-keying using Adobe After Effects or Nuke~\cite{AdobeAE,Nuke}.

To add motion to these characters, we tried several image-to-video models. However, the outputs featured extraneous motions that did not match the sharp, minimal, highly-focused motion style of 2D animation we desired. Instead, we used three general workflows that we found most successful, with AI portions implemented in ComfyUI~\cite{ComfyUI}.

\paragraph{AI puppets.} For simple shots, in which a character had only a single primary key pose, we used an artist sketch as input to ControlNet to generate an inked-and-painted hero frame. Then SegmentAnything~\cite{kirillov2023segment} separated that frame into body parts. The resulting layers were imported into Adobe Character Animator~\cite{AdobeCh} for pose animation and audio lip-sync. We removed the generated mouth shape using Photoshop~\cite{Photoshop}, and replaced it with a library of hand-drawn shapes using Character Animator's lip-sync tools.

\begin{figure}[h]
    \centering
    \includegraphics[width=0.66\linewidth]{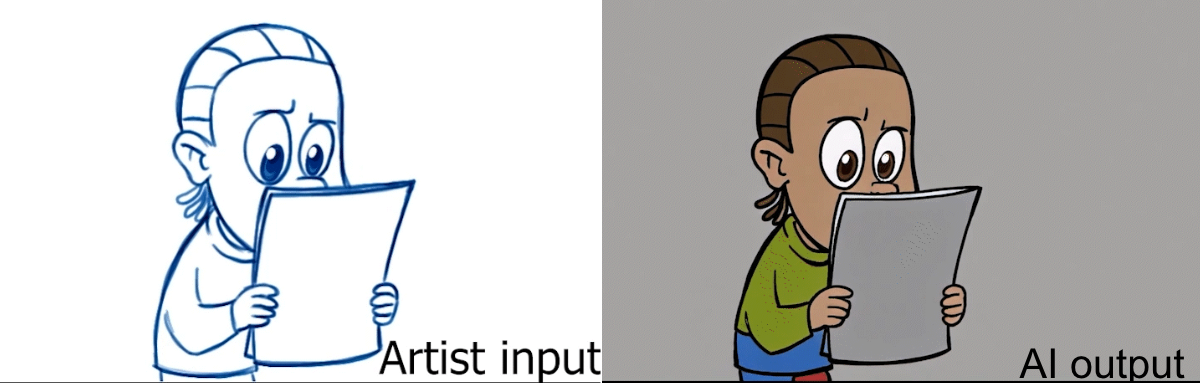}
    \caption{A sample frame of input and output for the character LUTHER from our AI ink and paint workflow.}
    \Description{A sample frame of input and output for the character LUTHER from our AI ink and paint workflow.}
    \label{fig:vid2vid}
\end{figure}

\paragraph{AI ink and paint.} For shots with more dramatic pose variations, animators drew pencil artwork per-frame, and we used AnimateDiff~\cite{guo2023animatediff} with ControlNet to generate consistent ``ink and paint'' (see Figure~\ref{fig:vid2vid}).  To ensure the ink lines matched the inputs, we use Canny-edge or sketch ControlNets, as well as depth ControlNet. Color deviations and edge noise in the output were corrected using DaVinci Resolve and Nuke~\cite{DaVinci,Nuke}.

\paragraph{SDEdit} Finally, we used a novel workflow for our radioactive FIREFLY, who is supposed to be a little crazy. We input frames from a smooth animation cycle into our SDXL model using an SDedit workflow~\cite{meng2022sdedit}. \todo{what noise amounts worked best?} This produced outputs which were similar, but not identical to the input frames (see Appendix). This randomness gave the FIREFLY a sense of constant, erratic motion.

\section{Conclusion}

AI workflows enabled animators to focus their time and effort on story, design and characterization, and less on tedious aspects of animation such as ink and paint. However, substantial manual cleanup was required to reach our target quality bar, and we fine-tuned our models repeatedly with different inputs to gain better intuitions. The ink and paint workflow worked best with very clean input lines; cleaner than our animators might normally produce for a pencil test. Nonetheless, diffusion models exceeded our initial expectations, enabling us to complete about 50 shots with only one full-time and two part-time animators over a production period of approximately eight weeks. We expect results to improve using base models trained or fine-tuned with more cartoon and animation samples, and with new conditioning strategies to constrain critical elements of character design such as proportion, line weight, and color palette.

\begin{acks}
Thanks to the Not-So-Supervillains creative and production crew: Jason Mayland, Kelly McNutt, Woei Lee, Gabe Liu, David Andrews, Eloise Fassler, Kavitha Gopalakrishnan, HOPR. Voice talent: Kory Mathewson, Warren Reid, Kiara Lee, Andru Anderson. Brain trust: Kory Mathewson, Cassidy Curtis, Alonso Martinez, Anna Kipnis. PM: Thomas Iljic.
\end{acks}

\bibliographystyle{ACM-Reference-Format}
\bibliography{main}


\begin{thebibliography}{14}


\ifx \showCODEN    \undefined \def \showCODEN     #1{\unskip}     \fi
\ifx \showDOI      \undefined \def \showDOI       #1{#1}\fi
\ifx \showISBNx    \undefined \def \showISBNx     #1{\unskip}     \fi
\ifx \showISBNxiii \undefined \def \showISBNxiii  #1{\unskip}     \fi
\ifx \showISSN     \undefined \def \showISSN      #1{\unskip}     \fi
\ifx \showLCCN     \undefined \def \showLCCN      #1{\unskip}     \fi
\ifx \shownote     \undefined \def \shownote      #1{#1}          \fi
\ifx \showarticletitle \undefined \def \showarticletitle #1{#1}   \fi
\ifx \showURL      \undefined \def \showURL       {\relax}        \fi
\providecommand\bibfield[2]{#2}
\providecommand\bibinfo[2]{#2}
\providecommand\natexlab[1]{#1}
\providecommand\showeprint[2][]{arXiv:#2}

\bibitem[DaV(2024)]%
        {DaVinci}
Blackmagic Design \bibinfo{year}{2024}\natexlab{}.
\newblock \bibinfo{booktitle}{\emph{DaVinci Resolve 18 | Blackmagic Design}}.
\newblock Blackmagic Design.
\newblock
\urldef\tempurl%
\url{https://www.blackmagicdesign.com/products/davinciresolve}
\showURL{%
Retrieved April 10, 2024 from \tempurl}


\bibitem[Ado(2024a)]%
        {AdobeCh}
Adobe \bibinfo{year}{2024}\natexlab{a}.
\newblock \bibinfo{booktitle}{\emph{Motion Capture Animation Software | Adobe
  Character Animator}}.
\newblock Adobe.
\newblock
\urldef\tempurl%
\url{https://www.adobe.com/products/character-animator.html}
\showURL{%
Retrieved April 10, 2024 from \tempurl}


\bibitem[Ado(2024b)]%
        {AdobeAE}
Adobe \bibinfo{year}{2024}\natexlab{b}.
\newblock \bibinfo{booktitle}{\emph{Motion Graphics Software | Adobe After
  Effects}}.
\newblock Adobe.
\newblock
\urldef\tempurl%
\url{https://www.adobe.com/products/aftereffects.html}
\showURL{%
Retrieved April 10, 2024 from \tempurl}


\bibitem[Nuk(2024)]%
        {Nuke}
Foundry \bibinfo{year}{2024}\natexlab{}.
\newblock \bibinfo{booktitle}{\emph{Nuke | VFX and Film Editing Software}}.
\newblock Foundry.
\newblock
\urldef\tempurl%
\url{https://www.foundry.com/products/nuke-family/nuke}
\showURL{%
Retrieved April 23, 2024 from \tempurl}


\bibitem[Pho(2024)]%
        {Photoshop}
Adobe \bibinfo{year}{2024}\natexlab{}.
\newblock \bibinfo{booktitle}{\emph{Official Adobe Photoshop | Photo \& Design
  Software}}.
\newblock Adobe.
\newblock
\urldef\tempurl%
\url{https://www.adobe.com/products/photoshop.html}
\showURL{%
Retrieved April 19, 2024 from \tempurl}


\bibitem[comfyanonymous(2024)]%
        {ComfyUI}
\bibfield{author}{\bibinfo{person}{comfyanonymous}.}
  \bibinfo{year}{2024}\natexlab{}.
\newblock \bibinfo{booktitle}{\emph{ComfyUI: The most powerful and modular
  stable diffusion GUI and backend}}.
\newblock GitHub.
\newblock
\urldef\tempurl%
\url{https://github.com/comfyanonymous/ComfyUI}
\showURL{%
Retrieved April 11, 2024 from \tempurl}


\bibitem[Guo et~al\mbox{.}(2024)]%
        {guo2023animatediff}
\bibfield{author}{\bibinfo{person}{Yuwei Guo}, \bibinfo{person}{Ceyuan Yang},
  \bibinfo{person}{Anyi Rao}, \bibinfo{person}{Zhengyang Liang},
  \bibinfo{person}{Yaohui Wang}, \bibinfo{person}{Yu Qiao},
  \bibinfo{person}{Maneesh Agrawala}, \bibinfo{person}{Dahua Lin}, {and}
  \bibinfo{person}{Bo Dai}.} \bibinfo{year}{2024}\natexlab{}.
\newblock \showarticletitle{AnimateDiff: Animate Your Personalized
  Text-to-Image Diffusion Models without Specific Tuning}.
\newblock \bibinfo{journal}{\emph{International Conference on Learning
  Representations}} (\bibinfo{year}{2024}).
\newblock


\bibitem[Hu et~al\mbox{.}(2021)]%
        {hu2021lora}
\bibfield{author}{\bibinfo{person}{Edward~J. Hu}, \bibinfo{person}{Yelong
  Shen}, \bibinfo{person}{Phillip Wallis}, \bibinfo{person}{Zeyuan Allen-Zhu},
  \bibinfo{person}{Yuanzhi Li}, \bibinfo{person}{Shean Wang},
  \bibinfo{person}{Lu Wang}, {and} \bibinfo{person}{Weizhu Chen}.}
  \bibinfo{year}{2021}\natexlab{}.
\newblock \bibinfo{title}{LoRA: Low-Rank Adaptation of Large Language Models}.
\newblock
\newblock
\showeprint[arxiv]{2106.09685}~[cs.CL]


\bibitem[Kirillov et~al\mbox{.}(2023)]%
        {kirillov2023segment}
\bibfield{author}{\bibinfo{person}{Alexander Kirillov}, \bibinfo{person}{Eric
  Mintun}, \bibinfo{person}{Nikhila Ravi}, \bibinfo{person}{Hanzi Mao},
  \bibinfo{person}{Chloe Rolland}, \bibinfo{person}{Laura Gustafson},
  \bibinfo{person}{Tete Xiao}, \bibinfo{person}{Spencer Whitehead},
  \bibinfo{person}{Alexander~C Berg}, \bibinfo{person}{Wan-Yen Lo},
  {et~al\mbox{.}}} \bibinfo{year}{2023}\natexlab{}.
\newblock \showarticletitle{Segment anything}. In
  \bibinfo{booktitle}{\emph{Proceedings of the IEEE/CVF International
  Conference on Computer Vision}}. \bibinfo{pages}{4015--4026}.
\newblock


\bibitem[Maas et~al\mbox{.}(2024)]%
        {Showrunner}
\bibfield{author}{\bibinfo{person}{Carey Maas}, \bibinfo{person}{Saatchi
  Wheeler}, \bibinfo{person}{Shamash Billington}, {et~al\mbox{.}}}
  \bibinfo{year}{2024}\natexlab{}.
\newblock \bibinfo{booktitle}{\emph{To infinity and beyond: Show-1 and
  Showrunner agents in multi-agent simulations}}.
\newblock FableStudio.
\newblock
\urldef\tempurl%
\url{https://fablestudio.github.io/showrunner-agents/}
\showURL{%
Retrieved April 19, 2024 from \tempurl}


\bibitem[Meng et~al\mbox{.}(2022)]%
        {meng2022sdedit}
\bibfield{author}{\bibinfo{person}{Chenlin Meng}, \bibinfo{person}{Yutong He},
  \bibinfo{person}{Yang Song}, \bibinfo{person}{Jiaming Song},
  \bibinfo{person}{Jiajun Wu}, \bibinfo{person}{Jun-Yan Zhu}, {and}
  \bibinfo{person}{Stefano Ermon}.} \bibinfo{year}{2022}\natexlab{}.
\newblock \showarticletitle{SDEdit: Guided Image Synthesis and Editing with
  Stochastic Differential Equations}. In
  \bibinfo{booktitle}{\emph{International Conference on Learning
  Representations}}.
\newblock


\bibitem[Podell et~al\mbox{.}(2023)]%
        {podell2023sdxl}
\bibfield{author}{\bibinfo{person}{Dustin Podell}, \bibinfo{person}{Zion
  English}, \bibinfo{person}{Kyle Lacey}, \bibinfo{person}{Andreas Blattmann},
  \bibinfo{person}{Tim Dockhorn}, \bibinfo{person}{Jonas Müller},
  \bibinfo{person}{Joe Penna}, {and} \bibinfo{person}{Robin Rombach}.}
  \bibinfo{year}{2023}\natexlab{}.
\newblock \bibinfo{title}{SDXL: Improving Latent Diffusion Models for
  High-Resolution Image Synthesis}.
\newblock
\newblock
\showeprint[arxiv]{2307.01952}~[cs.CV]


\bibitem[Ruiz et~al\mbox{.}(2023)]%
        {ruiz2023dreambooth}
\bibfield{author}{\bibinfo{person}{Nataniel Ruiz}, \bibinfo{person}{Yuanzhen
  Li}, \bibinfo{person}{Varun Jampani}, \bibinfo{person}{Yael Pritch},
  \bibinfo{person}{Michael Rubinstein}, {and} \bibinfo{person}{Kfir Aberman}.}
  \bibinfo{year}{2023}\natexlab{}.
\newblock \showarticletitle{Dreambooth: Fine tuning text-to-image diffusion
  models for subject-driven generation}. In
  \bibinfo{booktitle}{\emph{Proceedings of the IEEE/CVF Conference on Computer
  Vision and Pattern Recognition}}. \bibinfo{pages}{22500--22510}.
\newblock


\bibitem[Zhang et~al\mbox{.}(2023)]%
        {zhang2023adding}
\bibfield{author}{\bibinfo{person}{Lvmin Zhang}, \bibinfo{person}{Anyi Rao},
  {and} \bibinfo{person}{Maneesh Agrawala}.} \bibinfo{year}{2023}\natexlab{}.
\newblock \showarticletitle{Adding conditional control to text-to-image
  diffusion models}. In \bibinfo{booktitle}{\emph{Proceedings of the IEEE/CVF
  International Conference on Computer Vision}}. \bibinfo{pages}{3836--3847}.
\newblock


\end{thebibliography}

\appendix
\pagebreak
\pagenumbering{arabic}
\renewcommand*{\thepage}{A\arabic{page}}
\renewcommand\thefigure{\thesection.\arabic{figure}}    

\section{Supplementary Material}
\setcounter{figure}{0}    

\begin{figure}[hb]
    \centering
    \includegraphics[width=0.45\linewidth]{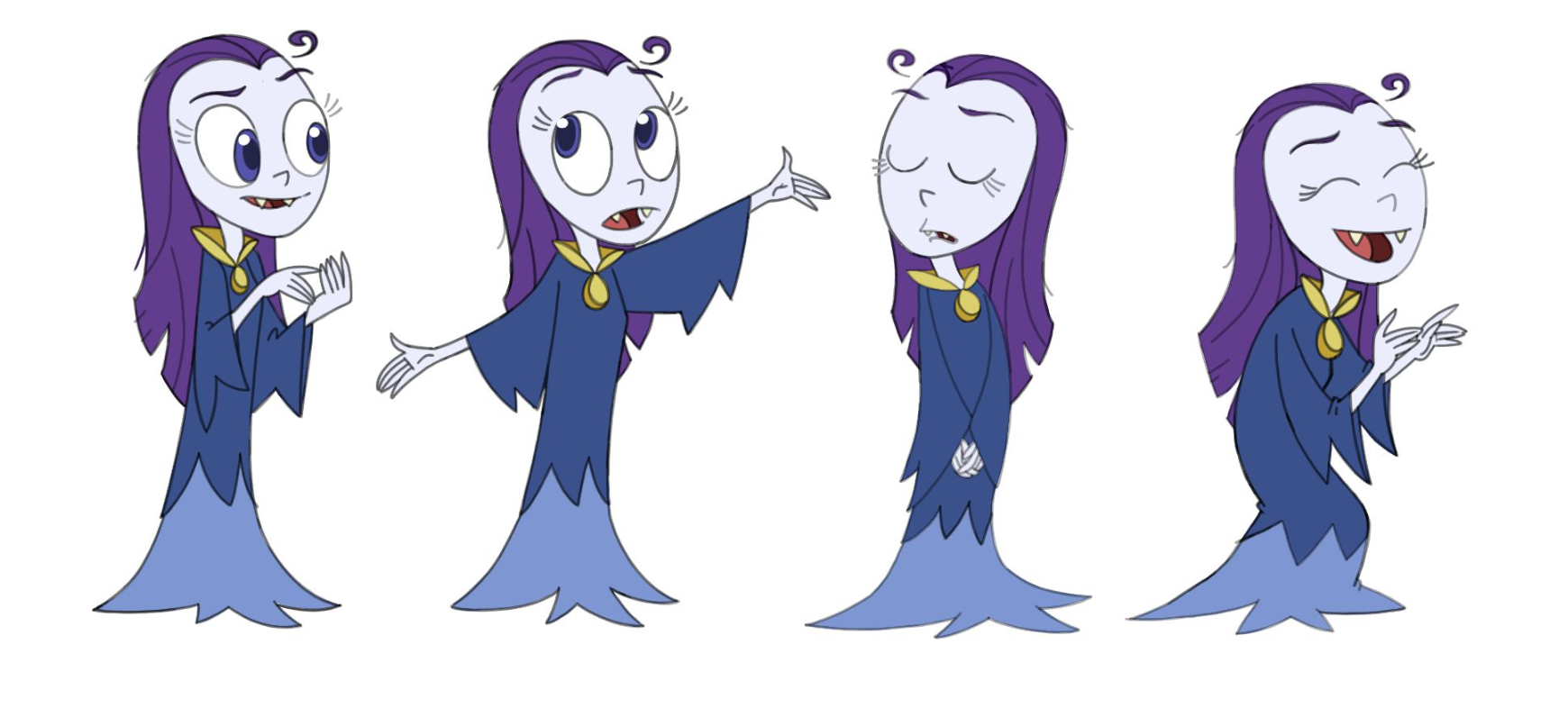}
    \includegraphics[width=0.2\linewidth]{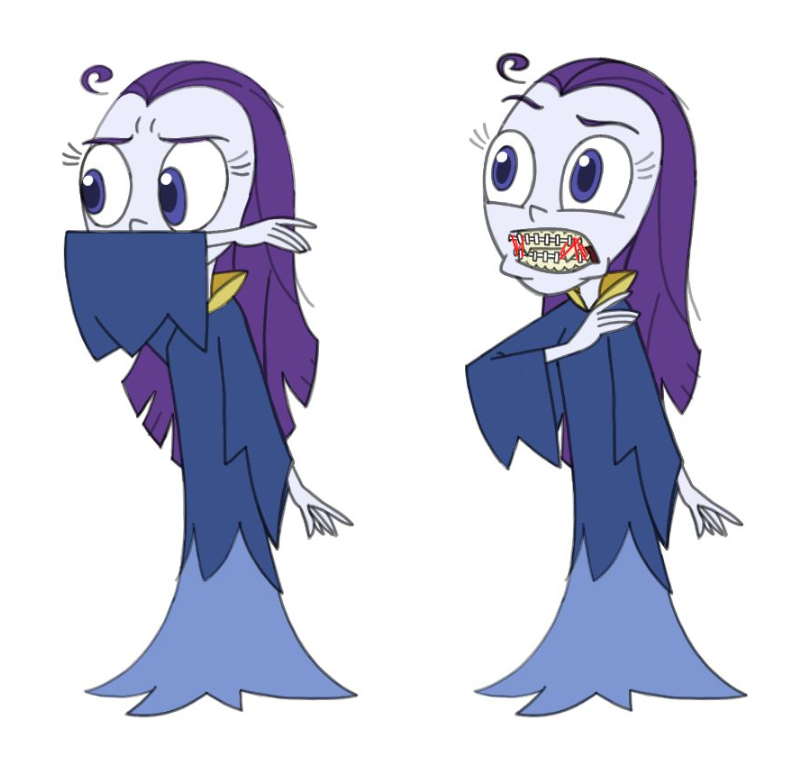}
    \includegraphics[width=0.3\linewidth]{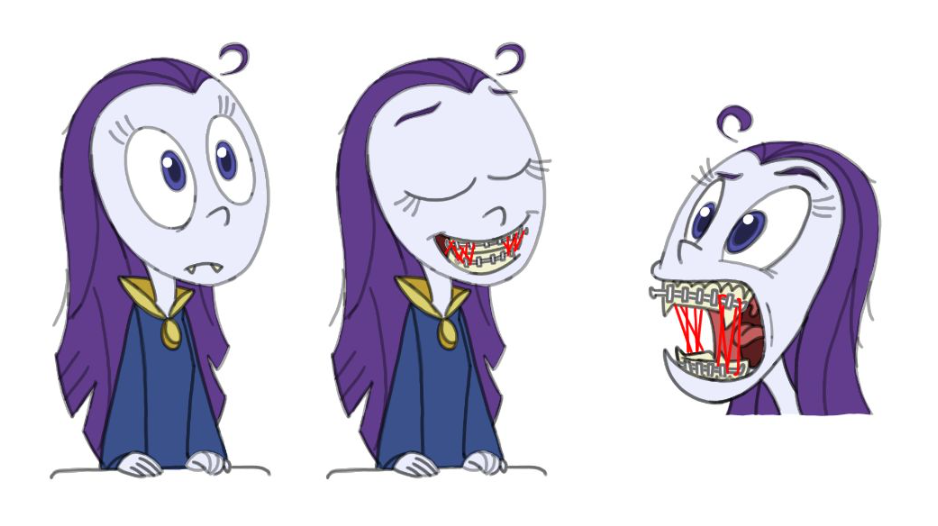}
    \caption{Hand-drawn illustrations used to fine-tune SDXL for the character VERNA. Character design by Kelly McNutt.}
    \Description{Nine hand-drawn illustrations used to fine-tune SDXL for the character VERNA.}
    \label{fig:verna_training}
\end{figure}

\begin{figure}[hb]
    \centering
    \includegraphics[width=0.45\linewidth]{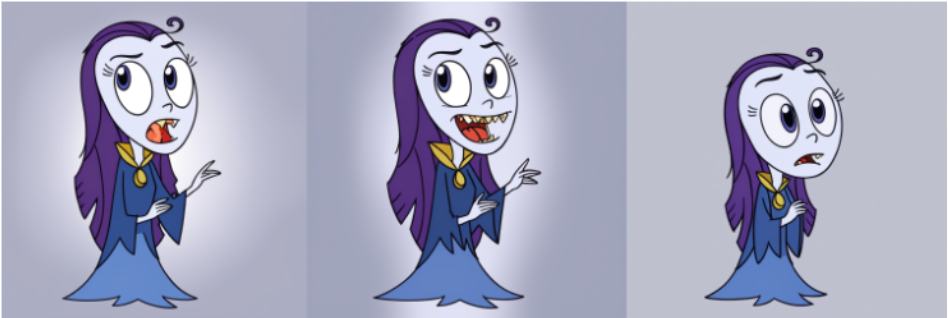}
    \includegraphics[width=0.45\linewidth]{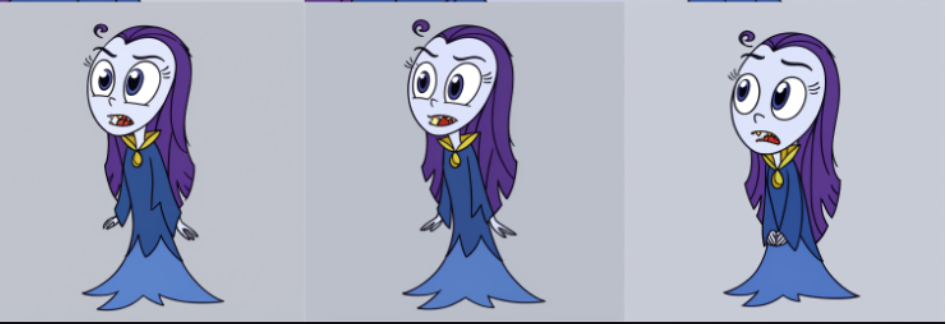}
    \caption{Sample text-to-image outputs for the VERNA DreamBooth model with no additional conditioning. Note the proportions of her head to body and eyes to face are inconsistent across different random seeds. However, other characteristics like the curl of her hair or the shape of her dress and hands are well-preserved across generations.}
    \Description{Six text-to-image model outputs for the VERNA model without conditioning, illustrating inconsistent proportions.}
    \label{fig:verna_t2i}
\end{figure}

\begin{figure}[hb]
    \centering
    \includegraphics[width=0.65\linewidth]{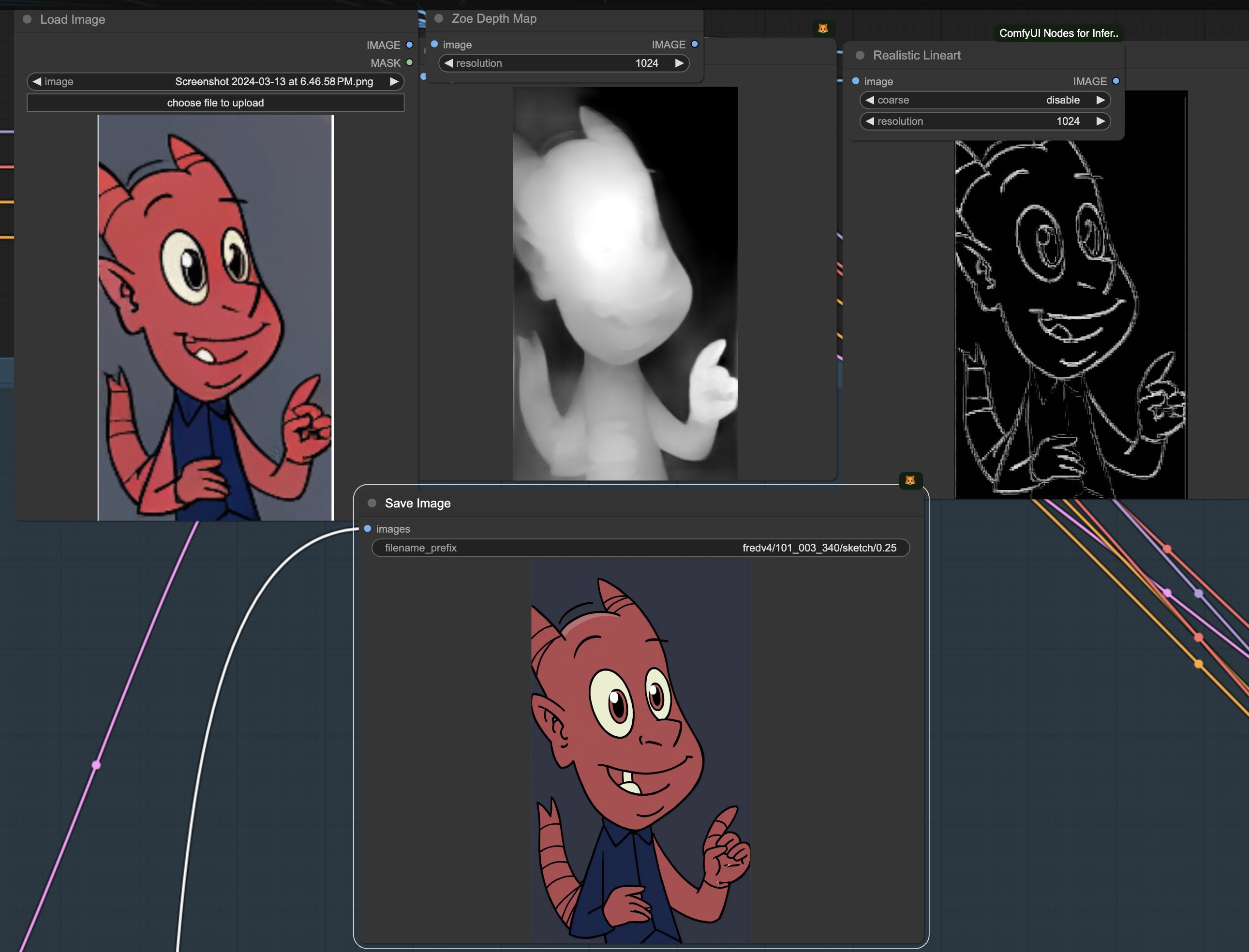}
    \caption{Part of a ComfyUI workflow for FRED MID-EVIL using depth and edge ControlNets.}
    \Description{Part of a ComfyUI workflow for FRED MID-EVIL using depth and edge ControlNets.}
    \label{fig:fred_comfyui}
\end{figure}

\begin{figure}[hb]
    \centering
    \includegraphics[width=\linewidth]{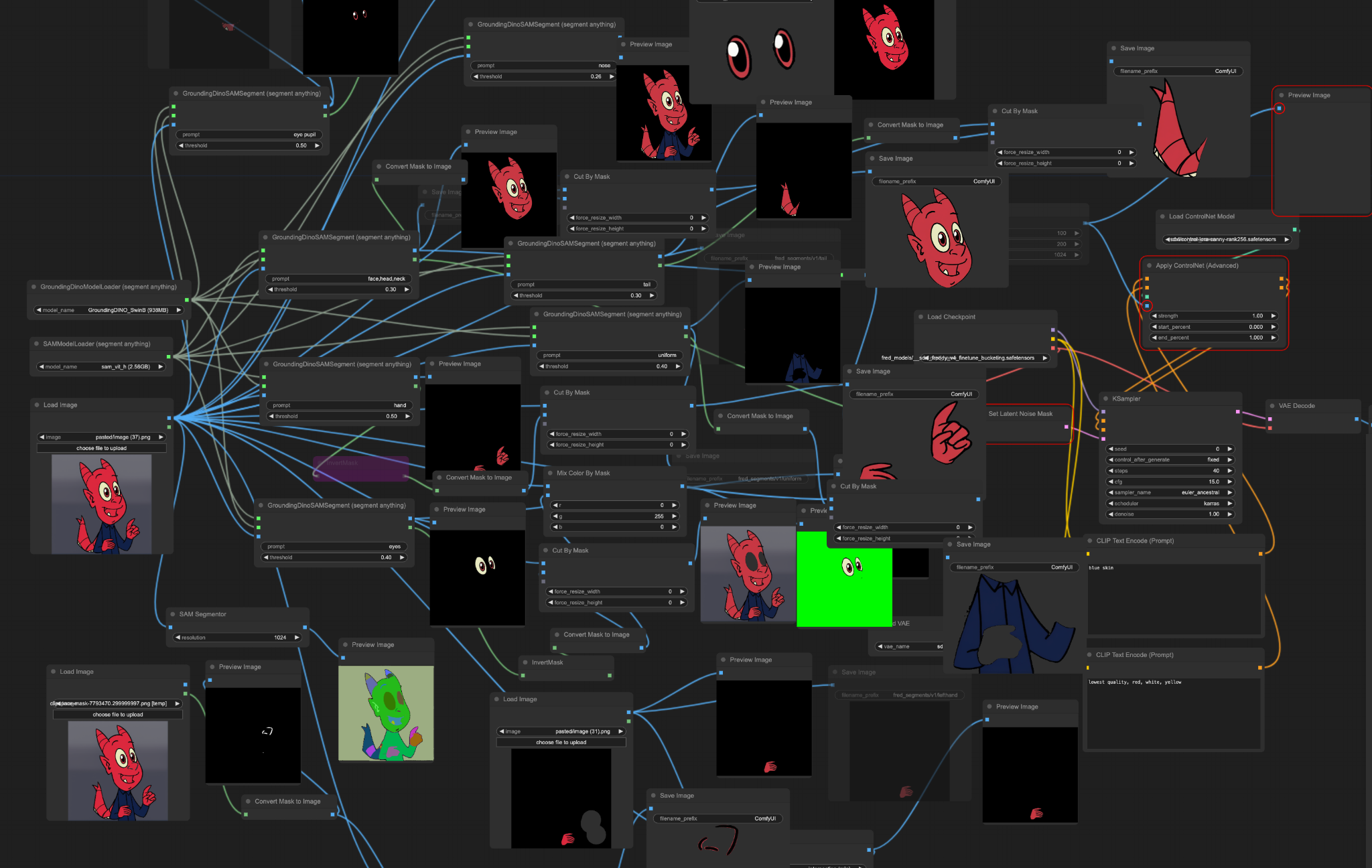}
    \caption{Part of a ComfyUI workflow using SegmentAnything to split FRED into layers.}
    \Description{Part of a ComfyUI workflow using SegmentAnything to split FRED into layers.}
    \label{fig:segmentanything}
\end{figure}

\begin{figure}[hb]
    \centering
    \includegraphics[width=\linewidth]{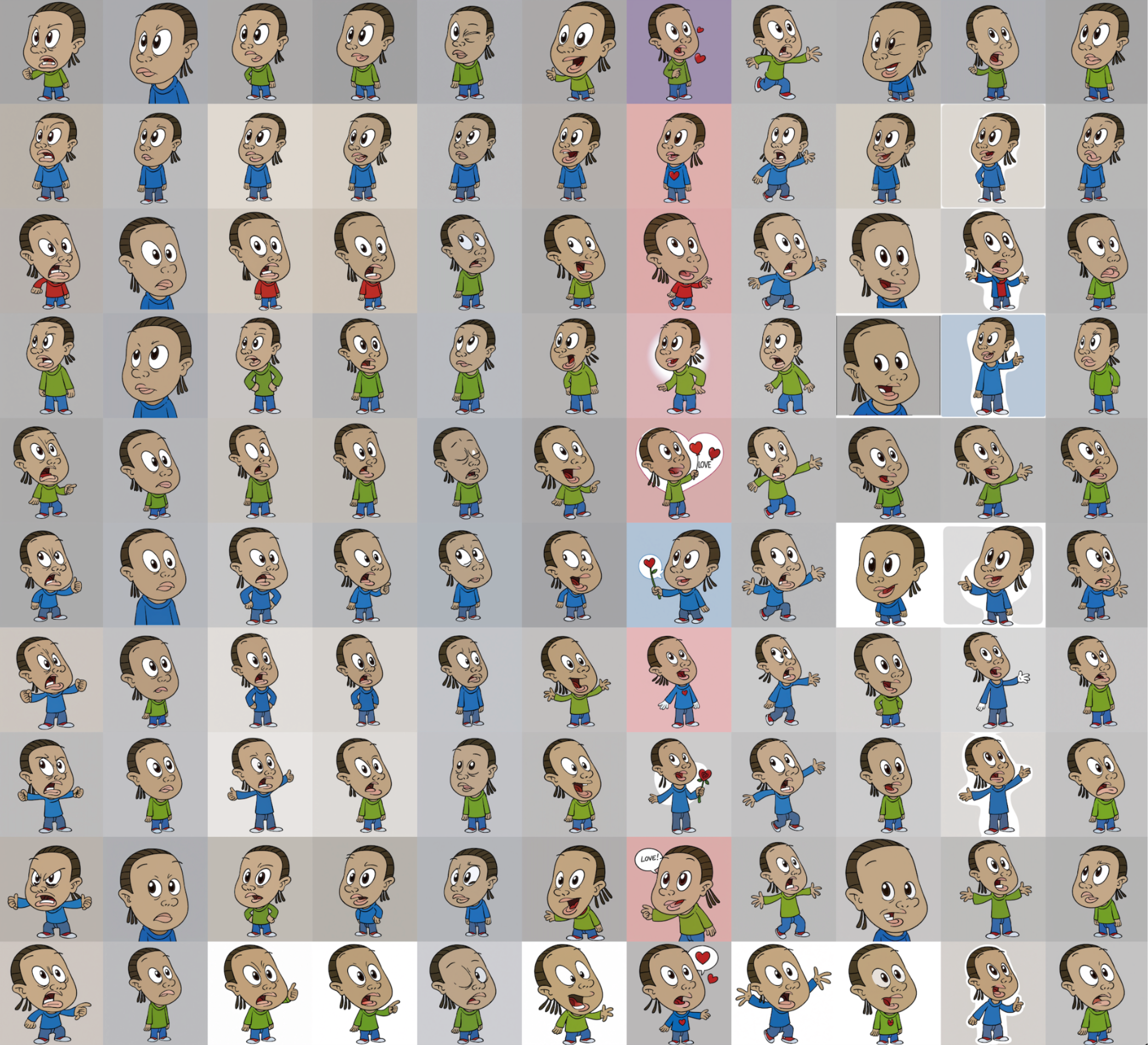}
    \caption{Sample generations of the LUTHER SPARKS model. Each row uses a different random seed, each column uses a prompt emphasizing a different emotion. From left to right: angry, confused, defiant, devious, sad, laughing, lovestruck, scared, smirking, proud, disgusted. While generated images sometimes convey emotions using body language, facial expressions were more limited.}
    \Description{Sample generations from the LUTHER SPARKS model, showing improved consistency and predictability when using ControlNet.}
    \label{fig:luther_samples}
\end{figure}

\begin{figure}[hb]
    \centering
    \includegraphics[width=0.8\linewidth]{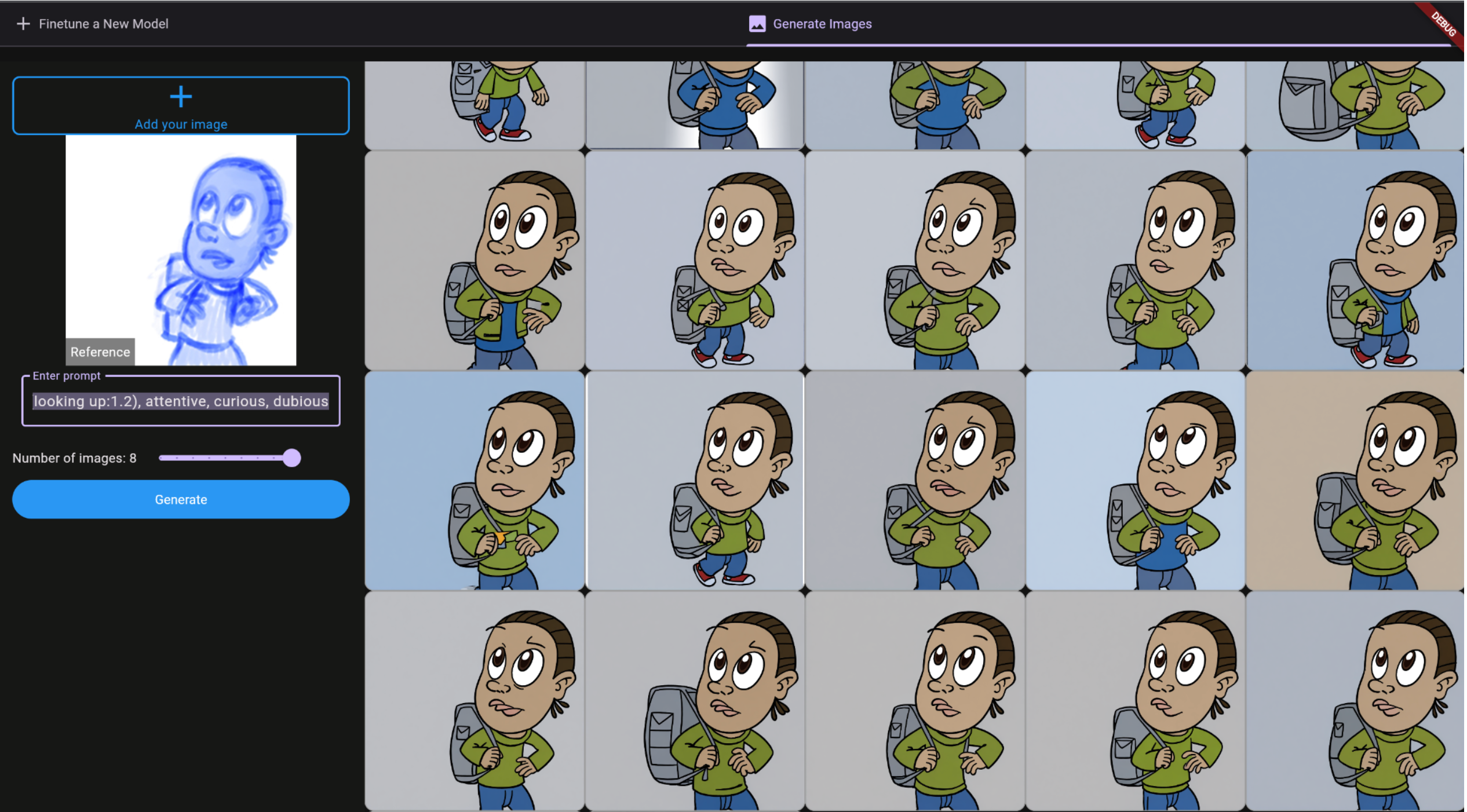}
    \caption{Sample generations of the LUTHER SPARKS model. Results became more consistent and predictable using ControlNet depth and/or edge conditioning, here shown using rough storyboard art by Kelly McNutt as input.}
    \Description{Sample generations from the LUTHER SPARKS model, showing improved consistency and predictability when using ControlNet.}
    \label{fig:luther_controlnet}
\end{figure}

\begin{figure}[hb]
    \centering
    \includegraphics[width=0.8\linewidth]{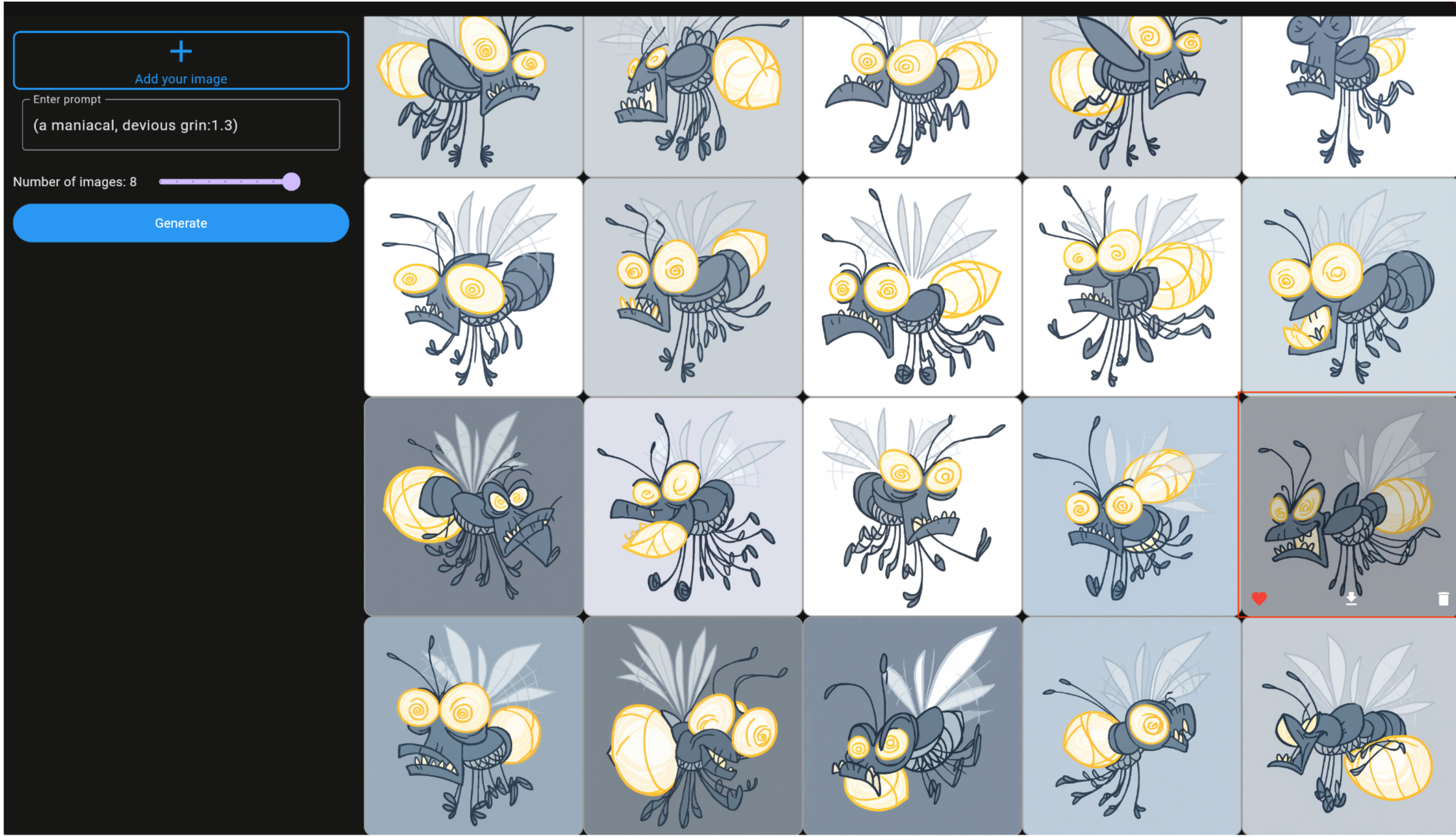}
    \caption{Sample generations from the FIREFLY model. Dreambooth training produced less consistent results for non-human characters, likely due to a domain gap with the diversity of the SDXL model's training data. We took advantage of this inconsistency to amplify this character's erratic and unpredictable nature!}
    \Description{Sample generations from the FIREFLY model, illustrating less consistent results for non-human characters.}
    \label{fig:firefly_samples}
\end{figure}


\end{document}